\documentclass[aps,preprint,amsmath,amsfonts,amssymb,showpacs,
showkeys]{revtex4}

\usepackage{bm}

\begin{document}

\title{Decay coupling constants sum rules for dibaryon octet into two baryon octets with $\bm{\lambda_8}$ first order SU(3) symmetry breaking}

\author{E. N.~Polanco-Eu\'an}

\email[]{elias.polanco@mda.cinvestav.mx}

\affiliation{ Departamento de F\'{\i}sica Aplicada.\\ Centro de
Investigaci\'on y de Estudios Avanzados del IPN.\\ Unidad
M\'erida.\\ A.P. 73, Cordemex.\\ M\'erida, Yucat\'an, 97310.
MEXICO. }

\author{V.~Gupta}

\email[]{virendra@mda.cinvestav.mx}

\affiliation{ Departamento de F\'{\i}sica Aplicada.\\ Centro de
Investigaci\'on y de Estudios Avanzados del IPN.\\ Unidad
M\'erida.\\ A.P. 73, Cordemex.\\ M\'erida, Yucat\'an, 97310.
MEXICO. }

\author{G.~S\'anchez-Col\'on}

\email[]{gabriel.sanchez@cinvestav.mx}

\affiliation{ Departamento de F\'{\i}sica Aplicada.\\ Centro de
Investigaci\'on y de Estudios Avanzados del IPN.\\ Unidad
M\'erida.\\ A.P. 73, Cordemex.\\ M\'erida, Yucat\'an, 97310.
MEXICO. }

\author{B. A.~Bambah}

\email[]{bbsp@uohyd.ernet.in}

\affiliation{ School of Physics.\\ University of Hyderabad.\\
Hyderabad, A.P. 500 046, India. }

\date{\today}

\begin{abstract}

The SU(3) octet states with baryon number $B=2$, hexaquark dibaryons, are considered. Decay coupling constants sum rules for dibaryon octet into two ordinary baryon octets with $\lambda_8$ first order SU(3) symmetry breaking are given. An SU(4) extension of the analysis is commented upon. Possibilities for the experimental observation of multibaryon and anti-multibaryon states are pointed out.

\end{abstract}

\pacs{14.20.Pt, 13.30.-a}

\keywords{dibaryon, sextaquarks, sum rule}

\maketitle

\section{\label{introduction}Introduction.}

The first hexaquark or sextaquark state with baryon number $B=2$, a dibaryon, was proposed in 1964~\cite{dyson64}. Hexaquarks received more attention after the proposal of the H-dibaryon ($uuddss$) by Jaffe~\cite{jaffe77}. The presence of hexaquark states have also been conjectured in models with colour flux tube (string) confinement,~\cite{nambu, bambah, Vijande}. The theoretical case for the dibaryon (H-dibaryon) has recently been made stronger by lattice-QCD calculations in the SU(3) flavor limit by the NPLQCD and HALQCD collaborations~\cite{lattice1,lattice2}. The presence of multiquark states including the hexaquark in hot-dense matter is predicted in certain statistical~\cite{stat} and holographic models~\cite{Burikham}, and since strangeness enhancement is associated with the QGP, hexaquarks with strangeness may have an observable signature in the upcoming heavy ion experiments at LHC. More recently Bashkanov {\it et.\ al}~\cite{bashkanov13} have suggested novel dibaryon states and their possible experimental signatures. These include exclusive measurements of the $n\,p\rightarrow n\,p\,\pi^0\pi^0$ cross section at the WASA detector in COSY, Julich, which have shown the presence of a $d^*(2348)$ resonance that can be interpreted as a dibaryon state~\cite{Julich}. Recently in Ref.~\cite{gupta15}, it was noted that the deuteron, a stable dibaryon, belongs to the $B=2$, $\bm{10^*}$ representation in the SU(3) $\bm{8}\otimes\bm{8}$ decomposition of two $J^P=(1/2)^+$ baryon octets (denoted here as $B_{\bm{8}}$) and sum rules for decays of $\bm{10^*}$ and $\bm{10}$ dibaryons decuplets (denoted $D_{\bm{10^*,10}}$) into an antisymmetric final state of two octet baryons, $D_{\bm{10^*,10}}\to B_{\bm{8}}+B_{\bm{8}}$, were given. A recent high statistics search in inclusive Upsilon decays was done for H-dibaryons in the symmetric SU(3) state at the Belle detector of KEKB $e^+e^−$ collider~\cite{kim12}, which put stringent constraints on the H-dibaryon mass and clearly showed that if a H-dibaryon exists in the mass range $\le 2\,m_{\Lambda}$, then it must be different from a deuteron. M.~Oka has also proposed the possibility of several dibaryon resonances examining flavor SU(3) octet dibaryon systems employing the quark cluster model~\cite{oka88}. Thus, it is important to study the various decay channels of a dibaryon belonging to the dibaryon octet.

In this paper we present sum rules for decays of the dibaryon SU(3) octet (denoted $D_{\bm{8}}$) into two ordinary $J^P=(1/2)^+$ baryon octets with $\lambda_8$ first order SU(3) symmetry breaking. This additional study of the symmetry breaking mechanism at SU(3) level sets the stage for how to treat the breaking of SU(4) to get reasonable estimates for couplings of charmed hexaquark states.

In Sec.~\ref{masses} the baryonic content of the $D_{\bm{8}}$ states and their expected masses are presented. In Sec.~\ref{sumrulesD8} decay sum rules in broken SU(3) are  determined. In Sec.~\ref{su4} the extension to the SU(4) case is considered.

Finally we discuss some open problems for multiquark states.

\section{\label{masses} Baryonic content and masses of dibaryon octet}

The $(Y,I)$ states (with hypercharge $Y$ and isospin $I$) of the dibaryon octet, denoted by $D_{\bm{8}}(Y,I)$ are given in Table~\ref{tablai}. The last column gives an estimate of the average mass of $D_{\bm{8}}(Y,I)$ depending on its baryonic content. From these we obtain:

\begin{equation}
3D_{\bm{8}}(0,0)+D_{\bm{8}}(0,1) = 9286.2{\rm MeV},
\end{equation}

\noindent
and

\begin{equation}
2[D_{\bm{8}}(1,1/2)+D_{\bm{8}}(-1,1/2)] =9132.0{\rm MeV}.
\end{equation}

\noindent
It is remarkable how well these estimate satisfy the first order Gell Mann-Okubo mass formula for the octet~\cite{gellmann61,okubo62}.

\section{\label{sumrulesD8}
Sum rules for $\bm{D_{8}\to B_8+B_8}$ decays with first order symmetry breaking of SU(3).}

We consider now dibaryon octet decays into a final state of two baryon octets: $D_{\bm{8}}\to B_{\bm{8}}+B_{\bm{8}}$. Following Ref.~\cite{carruthers} we write $B_{\bm{8}}$ and $D_{\bm{8}}$ matrices as:

\begin{equation}
B_{\bm{8}}=\left(
\begin{array}{ccc}
\frac{1}{\sqrt{6}}\Lambda+\frac{1}{\sqrt{2}}\Sigma^0 & \Sigma^+ & p \\
\Sigma^- & \frac{1}{\sqrt{6}}\Lambda-\frac{1}{\sqrt{2}}\Sigma^0 & n \\
-\Xi^- & \Xi^0 & -\sqrt{\frac{2}{3}} \Lambda
\end{array}
\right),
\end{equation}

\begin{equation}
D_{\bm{8}}=
\left(
\begin{array}{ccc}
\frac{1}{\sqrt{6}}D_{\bm{8}}(0,0,0)
+ \frac{1}{\sqrt{2}}D_{\bm{8}}(0,1,0) & D_{\bm{8}}(0,1,+1) &
D_{\bm{8}}(1,1/2,+1/2) \\
D_{\bm{8}}(0,1,-1) &
\frac{1}{\sqrt{6}}D_{\bm{8}}(0,0,0)-\frac{1}{\sqrt{2}}D_{\bm{8}}(0,1,0)
& D_{\bm{8}}(1,1/2,-1/2) \\
- D_{\bm{8}}(-1,1/2,-1/2) &
D_{\bm{8}}(-1,1/2,+1/2) & -\sqrt{\frac{2}{3}} D_{\bm{8}}(0,0,0)
\end{array}
\right),
\end{equation}

\noindent
where the $(Y,I,I_3)$ states of $D_{\bm{8}}$ are denoted by $D_{\bm{8}}(Y,I,I_3)$ with $I_3$ the isospin third component.

In general, for exact SU(3) the Yukawa interaction for $D_{\bm{8}}\to B_{\bm{8}}+B_{\bm{8}}$ ($\bm{8\to 8\otimes 8}$) is characterized by two parameters, $g^0_{\bm{8}}$  and $g^0_{\bm{8'}}$, since there are two octets in the decomposition:

\begin{equation}
\bm{8\otimes 8}=\bm{1}\oplus\bm{8}\oplus \bm{8'}\oplus \bm{10}\oplus\bm{10^{*}}\oplus \bm{27}.
\label{8x8}
\end{equation}

\noindent
The $\bm{1}$, $\bm{8}$, and $\bm{27}$ are symmetric under the interchange of the original two octets and $\bm{8'}$, $\bm{10}$, and $\bm{10^{*}}$ are antisymmetric.

With $\lambda_8$ first order breaking of SU(3) we expect eight additional parameters in addition to $g^0_{\bm{8}}$ and $g^0_{\bm{8'}}$ since the one dimensional representation $\bm{1}$ appears eight times in the decomposition $\bm{8\otimes 8\otimes 8\otimes 8}$, four corresponding to couplings with flavor antisymmetric final state representations and four corresponding to couplings with flavor symmetric final state representations.

In this work we will analyze both cases, antisymmetric and symmetric final state, separately.

\subsection{\label{d8A} Decays of octet dibaryon into $B_{\bm{8}}+B_{\bm{8}}$ antisymmetric final state.}

We consider decays of the dibaryon octet into a flavor antisymmetric final state of two baryon octets. As described above, in this case we have five parameters: $g^0_{\bm{8'}}$ from exact SU(3) and $g_k$ $(k=1,2,3,4)$ from $\lambda_8$ first order SU(3) symmetry breaking. The $D_{\bm{8}}\to B_{\bm{8}}+B_{\bm{8}}$ Yukawa couplings in terms of $g^0_{\bm{8'}}$ and $g_k$ $(k=1,2,3,4)$ are (for sake of clarity, we put a prime on the baryon matrix in the second octet  and $\overline{D}_{\bm{8}}=D^{\intercal}_{\bm{8}}$):

\begin{equation}
g^0_{\bm{8'}}\,{\rm Tr}\left\{\overline{D}_{\bm{8}}\left[B_{\bm{8}}B'_{\bm{8}}-B'_{\bm{8}}B_{\bm{8}}\right]\right\},
\label{1}
\end{equation}

\begin{equation}
g_1\,{\rm Tr}\left\{\overline{D}_{\bm{8}}\lambda_8\left[B_{\bm{8}}B'_{\bm{8}}-B'_{\bm{8}}B_{\bm{8}}\right]\right\},
\label{2}
\end{equation}

\begin{equation}
g_2\,{\rm Tr}\left\{\overline{D}_{\bm{8}}\left[B_{\bm{8}}\lambda_8\,B'_{\bm{8}}-B'_{\bm{8}}\lambda_8\,B_{\bm{8}}\right]\right\},
\label{3}
\end{equation}

\begin{equation}
g_3\,{\rm Tr}\left\{\overline{D}_{\bm{8}} \left[B_{\bm{8}}B'_{\bm{8}}-B'_{\bm{8}}B_{\bm{8}}\right]\lambda_8\right\},
\label{4}
\end{equation}

\begin{equation}
g_4\,\left\{ {\rm Tr}\left[\overline{D}_{\bm{8}}B_{\bm{8}}\right] {\rm Tr}\left[B'_{\bm{8}}\lambda_8\right]- {\rm Tr}\left[\overline{D}_{\bm{8_{\rm A}}}B'_{\bm{8}}\right] {\rm Tr}\left[B_{\bm{8}}\lambda_8\right]\right\}.
\label{5}
\end{equation}

There are eight independent decay coupling constants to be determined. We take these eight constants to be:

\begin{equation}
G[D_{\bm{8}}(1,1/2,+1/2)\to p\,{\Sigma^0}'\,],\quad G[D_{\bm{8}}(1,1/2,+1/2)\to p\,\Lambda'\,],
\label{34}
\end{equation}

\begin{equation}
G[D_{\bm{8}}(0,1,+1)\to p\,{\Xi^0}'\,],\ G[D_{\bm{8}}(0,1,+1)\to\Sigma^+\,\Lambda'\,],
\ G[D_{\bm{8}}(0,1,+1)\to\Sigma^+\,{\Sigma^0}'\,],
\label{35}
\end{equation}

\begin{equation}
G[D_{\bm{8}}(0,0,0)\to p\,{\Xi^-}'\,],
\label{36}
\end{equation}

\begin{equation}
G[D_{\bm{8}}(-1,1/2,+1/2)\to\Sigma^+\,{\Xi^-}'\,],\quad G[D_{\bm{8}}(-1,1/2,+1/2)\to\Lambda\,{\Xi^0}'\,].
\label{37}
\end{equation}

\noindent
The remaining decay coupling constants are given by:

\begin{eqnarray}
\lefteqn{
\sqrt{2}\,G[D_{\bm{8}}(1,1/2,+1/2)\to p\,{\Sigma^0}'\,] = -G[D_{\bm{8}}(1,1/2,+1/2)\to n\,{\Sigma^+}'\,] =
}
\nonumber\\
&= & G[D_{\bm{8}}(1,1/2,-1/2)\to p\,{\Sigma^-}'\,] = -\sqrt{2}\,G[D_{\bm{8}}(1,1/2,-1/2)\to n\,{\Sigma^0}'\,],
\label{D8Aothers1}
\end{eqnarray}

\begin{equation}
G[D_{\bm{8}}(1,1/2,+1/2)\to p\,\Lambda'\,]=G[D_{\bm{8}}(1,1/2,-1/2)\to n\,\Lambda'\,],
\label{D8Aothers2}
\end{equation}

\begin{equation}
G[D_{\bm{8}}(0,1,+1)\to\Sigma^+\,\Lambda'\,]= G[D_{\bm{8}}(0,1,0)\to\Sigma^0\,\Lambda'\,] = G[D_{\bm{8}}(0,1,-1)\to\Sigma^-\,\Lambda'\,],
\label{D8Aothers3}
\end{equation}

\begin{eqnarray}
G[D_{\bm{8}}(0,1,+1)\to p\,{\Xi^0}'\,]&=& \sqrt{2}\,G[D_{\bm{8}}(0,1,0)\to p\,{\Xi^-}'\,] = \sqrt{2}\,G[D_{\bm{8}}(0,1,0)\to n\,{\Xi^0}'\,] = 
\nonumber\\
&= &  G[D_{\bm{8}}(0,1,-1)\to n\,{\Xi^-}'\,],
\label{D8Aothers4}
\end{eqnarray}

\begin{equation}
G[D_{\bm{8}}(0,1,+1)\to\Sigma^+\,{\Sigma^0}'\,] = G[D_{\bm{8}}(0,1,0)\to\Sigma^+\,{\Sigma^-}'\,] = G[D_{\bm{8}}(0,1,-1)\to\Sigma^0\,{\Sigma^-}'\,],
\label{D8Aothers5}
\end{equation}

\begin{equation}
G[D_{\bm{8}}(0,0,0)\to n\,{\Xi^0}'\,] = - G[D_{\bm{8}}(0,0,0)\to p\,{\Xi^-}'\,],
\label{D8Aothers6}
\end{equation}

\begin{eqnarray}
\lefteqn{
G[D_{\bm{8}}(-1,1/2,+1/2)\to\Sigma^+\,{\Xi^-}'\,] = -\sqrt{2}\,G[D_{\bm{8}}(-1,1/2,+1/2)\to\Sigma^0\,{\Xi^0}'\,] =
}
\nonumber\\
&= &  \sqrt{2}\,G[D_{\bm{8}}(-1,1/2,-1/2)\to\Sigma^0\,{\Xi^-}'\,] = -G[D_{\bm{8}}(-1,1/2,-1/2)\to\Sigma^-\,{\Xi^0}'\,],
\label{D8Aothers7}
\end{eqnarray}

\begin{equation}
G[D_{\bm{8}}(-1,1/2,+1/2)\to\Lambda\,{\Xi^0}'\,] = G[D_{\bm{8}}(-1,1/2,-1/2)\to\Lambda\,{\Xi^-}'\,].
\label{D8Aothers8}
\end{equation}

The eight independent decay coupling constants (\ref{34})-(\ref{37}) are described by five parameters $g^0_{\bm{8'}}$ and $g_k$ $(k=1,2,3,4)$. Thus, three SU(3) broken sum rules between the Yukawa couplings may be deduced:

\begin{eqnarray}
\lefteqn{
4\,G[D_{\bm{8}}(1,1/2,+1/2)\to p\,{\Sigma^0}'\,]=2\sqrt{3}\,G[D_{\bm{8}}(0,0,0)\to p\,{\Xi^-}'\,]
}
\nonumber\\
&& \mbox{}-\sqrt{2}\,G[D_{\bm{8}}(0,1,+1)\to p\,{\Xi^0}'\,]+2\,G[D_{\bm{8}}(0,1,+1)\to\Sigma^+\,{\Sigma^0}'\,]
\nonumber\\
&&\mbox{}-2\sqrt{2}\,G[D_{\bm{8}}(-1,1/2,+1/2)\to\Sigma^+\,{\Xi^-}'\,],
\label{unoa}
\end{eqnarray}

\begin{eqnarray}
\lefteqn{
2\sqrt{6}\,G[D_{\bm{8}}(1,1/2,+1/2)\to p\,\Lambda'\,]=3\sqrt{6}\,G[D_{\bm{8}}(0,0,0)\to p\,{\Xi^-}'\,]
}
\nonumber\\
&& \mbox{}+5\,G[D_{\bm{8}}(0,1,+1)\to p\,{\Xi^0}'\,]-\sqrt{2}\,G[D_{\bm{8}}(0,1,+1)\to\Sigma^+\,{\Sigma^0}'\,]
\nonumber\\
&& \mbox{}+2\sqrt{6}\,G[D_{\bm{8}}(-1,1/2,+1/2)\to\Xi^0\,\Lambda'\,],
\label{dosa}
\end{eqnarray}

\begin{eqnarray}
\lefteqn{
\sqrt{6}\,G[\,D_{\bm{8}}(0,1,+1)\to\Sigma^+\,\Lambda'\,]=
}
\nonumber\\
&&= 2\,G[D_{\bm{8}}(0,1,+1)\to p\,{\Xi^0}'\,]-\sqrt{2}\,G[D_{\bm{8}}(0,1,+1)\to\Sigma^+\,{\Sigma^0}'\,]
\nonumber\\
&& \mbox{}+3\,G[D_{\bm{8}}(-1,1/2,+1/2)\to\Sigma^+\,{\Xi^-}'\,]+\sqrt{6}\,G[D_{\bm{8}}(-1,1/2,+1/2)\to\Xi^0\,\Lambda'\,],
\label{tresa}
\end{eqnarray}

\noindent
with identical relationships for $G[D_{\bm{8}}(Y,I,I_3)\to B'\,B\,] =-G[D_{\bm{8}}(Y,I,I_3)\to B\,B'\,]$.

\subsubsection{\label{su4} Comments on the extension to SU(4).}

Exact SU(4) symmetry is often used to make estimates of charmed baryon couplings. This gives rise to wrong results as SU(4) is badly broken. In this section we comment on the extension to SU(4) of the analysis of Sec.~\ref{d8A}~\cite{gupta76}.

The SU(4) group representation includes a new quantum number charm, which comes as a new quark. The baryons fit into the product:

\begin{equation}
\bm{4}\otimes \bm{4}\otimes \bm{4}=\bm{4^{*}}\oplus \bm{20'}\oplus \bm{20'}\oplus \bm{20},
\label{4.1}
\end{equation}

\noindent
the $J^P=(1/2)^+$ baryons are assigned to the $\bm{20'}$ representations and the $J^P=(3/2)^+$ resonances are placed in the $\bm{20}$.

In Sec.~\ref{d8A} the decays of the dibaryon octet $D_{\bm{8}}$ into an antisymmetric final state $B_{\bm{8}}+B_{\bm{8}}$ were studied. Correspondingly, for charmed dibaryons the decays of the 64-plet, $D_{\bm{64}}$, into an antisymmetric final state $B_{\bm{20'}}+B_{\bm{20'}}$ have to be considered, as can be seen from the decomposition:

\begin{equation}
\bm{20'}\otimes \bm{20'} = \bm{6}\oplus\bm{10}\oplus \bm{10^{*}}\oplus \bm{50}\oplus \bm{64_{\rm A}}\oplus \bm{64_{\rm S}}\oplus \bm{70}\oplus \bm{126}.
\label{4.2}
\end{equation}

\noindent
For exact SU(4) the Yukawa interaction is characterized by only one coupling constant, since the antisymmetrical $\bm{64_{\rm A}}$ representation appears only once in the decomposition of $\bm{20'\otimes 20'}$.

For SU(4) breaking it can be assumed, by using the spurion method, that the symmetry breaking operator transforms like the $Y=I=0$ component of the SU(4) $\bm{15}$ representation. Denote the symmetry breaking interaction as a spurion, $S_{\bm{15}}$, the $(Y,I)=(0,0)$ state of the SU(4) $\bm{15}$, and consider the reaction

\begin{equation}
S_{\bm{15}}+D_{\bm{64}}\to B_{\bm{20'}}+B_{\bm{20'}}
\label{4.3}
\end{equation}

\noindent
as SU(4) invariant. Consider the decomposition

\begin{equation}
\bm{15}\otimes \bm{64} = \bm{6}\oplus \bm{10}\oplus \bm{10^{*}}\oplus \bm{50}\oplus 3(\bm{64})\oplus\bm{70}\oplus \bm{70^{*}}\oplus \bm{126}\oplus \bm{126^{*}}\oplus \bm{300},
\label{4.4}
\end{equation}

\noindent
and the decomposition of $\bm{20'}\otimes \bm{20'}$, Eq.~(\ref{4.2}), where the representations $\bm{6}$, $\bm{50}$, $\bm{64_{\rm A}}$, and $\bm{70}$ are antisymmetric. For the two final baryon $\bm{20'}$ be in a flavor antisymmetric state, we find that there are six SU(4) breaking terms to first order, namely, $\langle\bm{6}|\bm{6}\rangle$, $\langle\bm{50}|\bm{50}\rangle$, $\langle\bm{64}|\bm{64_{\rm A}}\rangle$, $\langle\bm{64'}|\bm{64_{\rm A}}\rangle$, $\langle\bm{64''}|\bm{64_{\rm A}}\rangle$, and $\langle\bm{70}|\bm{70}\rangle$. Sum rules for SU(4) breaking along with exact SU(3) could be calculated at this first stage. We plan to present these elsewhere.

%

\subsection{\label{d8S} Decays of octet dibaryon into $B_{\bm{8}}+B_{\bm{8}}$ symmetric final state.}

Now we consider decays of the dibaryon octet $D_{\bm{8}}$ into two baryon octets flavor symmetric final state. In principle, with $\lambda_8$ first order breaking of SU(3) there are five additional parameters $g'_i$ ($i=1,2,3,4,5$) in addition to $g^0_{\bm{8}}$:

\begin{equation}
g^0_{\bm{8}}\,{\rm Tr}\left\{\overline{D}_{\bm{8}}\left[B_{\bm{8}}B'_{\bm{8}} + B'_{\bm{8}}B_{\bm{8}}\right]\right\},
\label{S01}
\end{equation}

\begin{equation}
g'_1\,{\rm Tr}\left\{\overline{D}_{\bm{8}}\lambda_8\left[B_{\bm{8}}B'_{\bm{8}} + B'_{\bm{8}}B_{\bm{8}}\right]\right\},
\label{S02}
\end{equation}

\begin{equation}
g'_2\,{\rm Tr}\left\{\overline{D}_{\bm{8}}\left[B_{\bm{8}}\lambda_8\,B'_{\bm{8}} + B'_{\bm{8}}\lambda_8\,B_{\bm{8}}\right]\right\},
\label{S03}
\end{equation}

\begin{equation}
g'_3\,{\rm Tr}\left\{\overline{D}_{\bm{8}} \left[B_{\bm{8}}B'_{\bm{8}} + B'_{\bm{8}}B_{\bm{8}}\right]\lambda_8\right\},
\label{S04}
\end{equation}

\begin{equation}
g'_4\,\left\{ {\rm Tr}\left[\overline{D}_{\bm{8}}B_{\bm{8}}\right] {\rm Tr}\left[B'_{\bm{8}}\lambda_8\right] + {\rm Tr}\left[\overline{D}_{\bm{8_{\rm A}}}B'_{\bm{8}}\right] {\rm Tr}\left[B_{\bm{8}}\lambda_8\right]\right\}.
\label{S05}
\end{equation}

\begin{equation}
g'_5\,{\rm Tr}\left[\overline{D}_{\bm{8}}\lambda_8\right] {\rm Tr}\left[B_{\bm{8}}B'_{\bm{8}}\right].
\label{S06}
\end{equation}

\noindent
However, the terms in Eqs.~(\ref{S01})-(\ref{S06}) are not linearly independent, there is the mathematical identity:

\begin{eqnarray}
\lefteqn{
{\rm Tr}\left\{\overline{D}_{\bm{8}}\lambda_8\left[B_{\bm{8}}B'_{\bm{8}} + B'_{\bm{8}}B_{\bm{8}}\right]\right\} + {\rm Tr}\left\{\overline{D}_{\bm{8}}\left[B_{\bm{8}}\lambda_8\,B'_{\bm{8}} + B'_{\bm{8}}\lambda_8\,B_{\bm{8}}\right]\right\}
}
\nonumber\\
&& + {\rm Tr}\left\{\overline{D}_{\bm{8}} \left[B_{\bm{8}}B'_{\bm{8}} + B'_{\bm{8}}B_{\bm{8}}\right]\lambda_8\right\} =
\nonumber\\
&& {\rm Tr}\left[\overline{D}_{\bm{8}}B_{\bm{8}}\right] {\rm Tr}\left[B'_{\bm{8}}\lambda_8\right] + {\rm Tr}\left[\overline{D}_{\bm{8}}B'_{\bm{8}}\right] {\rm Tr}\left[B_{\bm{8}}\lambda_8\right] + {\rm Tr}\left[\overline{D}_{\bm{8}}\lambda_8\right] {\rm Tr}\left[B_{\bm{8}}B'_{\bm{8}}\right].
\label{S7}
\end{eqnarray}

\noindent
Using~(\ref{S7}) to eliminate the $g'_2$ term we finally obtain:

\begin{equation}
g^0_{\bm{8}}\,{\rm Tr}\left\{\overline{D}_{\bm{8}}\left[B_{\bm{8}}B'_{\bm{8}} + B'_{\bm{8}}B_{\bm{8}}\right]\right\},
\label{S1}
\end{equation}

\begin{equation}
g_1\,{\rm Tr}\left\{\overline{D}_{\bm{8}}\lambda_8\left[B_{\bm{8}}B'_{\bm{8}} + B'_{\bm{8}}B_{\bm{8}}\right]\right\},
\label{S2}
\end{equation}

\begin{equation}
g_2\,{\rm Tr}\left\{\overline{D}_{\bm{8}} \left[B_{\bm{8}}B'_{\bm{8}} + B'_{\bm{8}}B_{\bm{8}}\right]\lambda_8\right\},
\label{S4}
\end{equation}

\begin{equation}
g_3\,\left\{ {\rm Tr}\left[\overline{D}_{\bm{8}}B_{\bm{8}}\right] {\rm Tr}\left[B'_{\bm{8}}\lambda_8\right] + {\rm Tr}\left[\overline{D}_{\bm{8_{\rm S}}}B'_{\bm{8}}\right] {\rm Tr}\left[B_{\bm{8}}\lambda_8\right]\right\}.
\label{S5}
\end{equation}

\begin{equation}
g_4\,{\rm Tr}\left[\overline{D}_{\bm{8}}\lambda_8\right] {\rm Tr}\left[B_{\bm{8}}B'_{\bm{8}}\right].
\label{S6}
\end{equation}

\noindent
As described at the beginning of this section, we have five parameters in this case: $g^0_{\bm{8}}$ from exact SU(3) and $g_i$ ($i=1,2,3,4$) from $\lambda_8$ first order SU(3) symmetry breaking.

There are nine independent coupling constants to be determined, we take these to be:

\begin{equation}
G[D_{\bm{8}}(1,1/2,+1/2)\to p\,{\Sigma^0}'\,],\quad G[D_{\bm{8}}(1,1/2,+1/2)\to p\,\Lambda'\,],
\label{S34}
\end{equation}

\begin{equation}
G[D_{\bm{8}}(0,1,+1)\to p\,{\Xi^0}'\,],\quad G[D_{\bm{8}}(0,1,+1)\to\,\Sigma^+\,\Lambda'\,],
\label{S35}
\end{equation}

\begin{equation}
G[D_{\bm{8}}(0,0,0)\to p\,{\Xi^-}'\,],\ G[D_{\bm{8}}(0,0,0)\to \Sigma^+{\Sigma^-}'\,],
\ G[D_{\bm{8}}(0,0,0)\to \Lambda{\Lambda}'\,],
\label{S36}
\end{equation}

\begin{equation}
G[D_{\bm{8}}(-1,1/2,+1/2)\to\Sigma^+\,{\Xi^-}'\,],\quad  G[D_{\bm{8}}(-1,1/2,+1/2)\to\Lambda\,{\Xi^0}'\,].
\label{S37}
\end{equation}

\noindent
The remaining decay coupling constants are given by the same relations, Eqs.~(\ref{D8Aothers1})-(\ref{D8Aothers8}), with $G[D_{\bm{8}}(0,1,+1)\to\Sigma^+\,{\Sigma^0}'\,] =0$ and $G[D_{\bm{8}}(0,0,0)\to \Sigma^+{\Sigma^-}'\,] = - G[D_{\bm{8}}(0,0,0)\to\Sigma^0{\Sigma^0}'\,]$.

The nine coupling constants (\ref{S34})-(\ref{S37}) are described by the five parameters $g^0_{\bm{8}}$ and $g_i$ ($i=1,2,3,4$). Thus, four SU(3) broken sum rules between the decay couplings constants may be deduced:

\begin{eqnarray}
\lefteqn{
\sqrt{6}\,G[D_{\bm{8}}(1,1/2,+1/2)\to p\,\Lambda'\,]=-\sqrt{2}\,G[D_{\bm{8}}(1,1/2,+1/2)\to p\,{\Sigma^0}'\,]
}
\nonumber\\
&& \mbox{}-\sqrt{6}\,G[D_{\bm{8}}(-1,1/2,+1/2)\to\Lambda\,{\Xi^{0}}'\,]+\,G[D_{\bm{8}}(-1,1/2,+1/2)\to\Sigma^+\, {\Xi^-}'\,],
\label{unoSa}
\end{eqnarray}

\begin{eqnarray}
\lefteqn{
\sqrt{6}\,G[D_{\bm{8}}(0,1,+1)\to\Sigma^+\,\Lambda'\,]=-2\,G[D_{\bm{8}}(0,1,+1)\to p\,{\Xi^{0}}'\,]
}
\nonumber\\
&& \mbox{}-\sqrt{6}\,G[D_{\bm{8}}(-1,1/2,+1/2)\to\Lambda\,{\Xi^{0}}'\,]+G[D_{\bm{8}}(-1,1/2,+1/2)\to\Sigma^+\, {\Xi^-}'\,],
\label{dosSa}
\end{eqnarray}

\begin{eqnarray}
\lefteqn{
\sqrt{6}\,G[D_{\bm{8}}(0,0,0)\to p\,{\Xi^-}'\,]=2\sqrt{2}\,G[D_{\bm{8}}(1,1/2,+1/2)\to p\,{\Sigma^0}'\,]
}
\nonumber\\
&& \mbox{}+G[D_{\bm{8}}(0,1,+1)\to p\,{\Xi^{0}}'\,]+\sqrt{6}\,G[D_{\bm{8}}(0,0,0)\to\Sigma^+\,{\Sigma^-}'\,]
\nonumber\\
&& \mbox{}+2\,G[D_{\bm{8}}(-1,1/2,+1/2)\to\Sigma^+\,{\Xi^-}'\,],
\label{tresSa}
\end{eqnarray}

\begin{eqnarray}
\lefteqn{
3\sqrt{3}\,G\,[D_{\bm{8}}(0,0,0)\to\Lambda\,\Lambda'\,]\,=-8\,G[D_{\bm{8}}(1,1/2,+1/2)\to p\,{\Sigma^0}'\,]
}
\nonumber\\
&& \mbox{}-2\sqrt{2}\,G[D_{\bm{8}}(0,1,+1)\to p\,{\Xi^{0}}'\,]-3\sqrt{3}\,G[D_{\bm{8}}(0,0,0)\to\Sigma^+\,{\Sigma^-}'\,]
\nonumber\\
&& \mbox{}-6\sqrt{3}\,G[D_{\bm{8}}(-1,1/2,+1/2)\to\Lambda\,{\Xi^0}'\,]-\sqrt{2}\,G[D_{\bm{8}}(-1,1/2,+1/2)\to\Sigma^+\,{\Xi^-}'\,].
\label{cuatroSa}
\end{eqnarray}

\noindent
With identical relationships for $G[D_{\bm{8}}(Y,I,I_3)\to B'\,B\,] = G[D_{\bm{8}}(Y,I,I_3)\to B\,B'\,]$.

\section{\label{conclusions}Concluding remarks}

Earlier, decays of sextaquark dibaryon decuplets ($\bm{10^*}$ and $\bm{10}$ representations) into a flavor antisymmetric final state of two ordinary baryon octets were considered~\cite{gupta15}. In this paper we have given sum rules with $\lambda_8$ first order SU(3) symmetry breaking for decays of dibaryon octet $D_{\bm{8}}$ into two ordinary baryon octets $B_{\bm{8}}+B_{\bm{8}}$. The sum rules given here and in earlier papers~\cite{gupta15,gupta64} can be applied into $B(\bm{8})+\bar{B}(\bm{8})$ with appropiate changes.

Very recently LHCb it has been reported that a multiquark state $Z(4430)^-$ (tetraquark) made of charm, anti-charm, down and anti-up quarks~\cite{LHCb} exists. This opens up a whole new area of multiquark states with charm and thus an SU(4) extension of our results is called for.

Apart from the heavy ion experiments, which we mentioned in the introduction, in our view $e^+e^-$ annihilation reaction is best suited to find multibaryon plus antimultibaryon pairs, since the signal for resonance will be a clear signal. However LHCb will also be a good place for observing these states.

\begin{acknowledgments}

V.~Gupta, E.~N.~Polanco-Eu\'an, and G.~S\'anchez-Col\'on would like to thank CONACyT (M\'exico) for partial support. B.~A. Bambah would like to acknowledge Ms. A.~Aiswarya for checking some calculations.

\end{acknowledgments}

\clearpage

\begin{table}

\caption{
The $(Y,I)$ states $D_{\bm{8}}(Y,I)$ in the $D_{\bm{8}}$ dibaryon octet that one obtains in the reduction $\bm{8}\otimes\bm{8}$ of two ordinary baryon octets $B_{\bm{8}}+B_{\bm{8}}$. For the SU(3) Clebsch-Gordan coefficients see Refs.~\cite{deswart, carruthers}. The average mass (in MeV) is given in the last column.
\label{tablai}}

\begin{ruledtabular}

\begin{tabular}{lcc}

$D_{\bm{8}}(Y,I)$ & $B_{\bm{8}}+B_{\bm{8}}$ & Average mass \\ 
\hline $D_{\bm{8}}(1,\frac{1}{2})$ & $\frac{1}{2}[ (\Sigma
N')_{\frac{1}{2}} + (N\Sigma')_{\frac{1}{2}} +
(N\Lambda')_{\frac{1}{2}} - (\Lambda N')_{\frac{1}{2}} ]$ &
2093.5 \\

$D_{\bm{8}}(0,1)$ & $\frac{1}{\sqrt{6}}[ 2(\Sigma\Sigma')_1 + (N\Xi')_1
- (\Xi N')_1 ]$ & 2514.9 \\

$D_{\bm{8}}(0,0)$ & $\frac{1}{\sqrt{2}}[ (N\Xi')_{0} + (\Xi N')_{0} ]$
& 2251.5 \\

$D_{\bm{8}}(-1,\frac{1}{2})$ & $\frac{1}{\sqrt{2}}[
(\Sigma\Xi')_{\frac{1}{2}} + (\Xi\Sigma')_{\frac{1}{2}} +
(\Lambda\Xi')_{\frac{1}{2}} - (\Xi\Lambda')_{\frac{1}{2}} ]$ &
2472.5 \\
 
\end{tabular}

\end{ruledtabular}

\end{table}

\end{document}